# Entangled electron-photon pair production by channel-exchange in high-energy Compton scattering

Basudev Nag Chowdhury[1*], Sanatan Chattopadhyay[1]

[1]*Department of Electronic Science, University of Calcutta, India*

*email address: basudevn@gmail.com

**Abstract**

The present work theoretically investigates the probability of generation of entangled electron-photon pair in high-energy Compton scattering of unpolarized electrons and photons due to scattering-channel-exchange mechanism. The study suggests that the scattering of unpolarized electrons and photons with nearly equal energy of the order of MeV at cross-channel can create entangled pair of up-spin electron and RCP photon or down-spin electron and LCP photon. The entanglement is quite strong exhibiting jump-concurrence for particular directions of scattering however not limited by the Kapitza-Dirac restriction. Such strong electron-photon entanglement is observed to be highly selective along the directions of scattering on the surface of two opposite cones at $45^0$ around the electron-detector axis and on the plane perpendicular to it, the latter one exhibiting maximal entanglement. The work predicts that spin-polarization entangled electron-photon pairs are highly possible to be found in practical experiments of high-energy Compton scattering with specific directional selectivity.

**Keywords:** Electron-photon entanglement; high-energy Compton scattering; channel of Compton scattering; Kapitza-Dirac effect.

Entanglement related phenomena have attracted the research community for their enormous technological applications in quantum information processing as well as in further exploring the physics of so-called 'spooky' behavior observed in nature across vastly different scales [1-3]. Numerous efforts are being made to study theoretically and experimentally the various aspects of quantum entanglement [4-13]. Although most of such experiments are carried out describing entangled-photons, in recent times the electron-photon entanglement has gained significant importance due to its possible applications in long-distance quantum teleportation and quantum repeater technologies. In this regard a few experimental reports have demonstrated the quantum entanglement phenomena between electron spin and photon frequency/polarization [14-16].

However, in contrast to the low-energy domain, very few reports are available on the entanglement behavior of high-energy particles. Recently, the possibility of helicity/polarization-entangled pair-production in high-energy scattering of spin-polarized particles has been studied theoretically, which claims maximally entangled electron-photon pair-production to be impossible in high-energy Compton scattering [17]. However, another contributory work, as a part of a general study on polarization dynamics in Compton scattering, has shown that scattering of electron and photon polarized in specified directions can generate a maximally entangled electron-photon pair [18]. Such work, based on Kapitza-Dirac principle of reflection of electrons from standing light waves [19], claims that Compton scattering of a vertically polarized photon and an electron with spinors tilted at $45^0$-up/down with photon propagation direction creates entangled pairs either of a left-circularly polarized photon with a $45^0$-up polarized electron or a right-circularly polarized photon with a $45^0$-down polarized electron. Such phenomena of entangled electron-photon pair production, however, are shown to be limited by two basic requirements: firstly, the incoming and outgoing photons must propagate in opposite directions with identical momentum value (i.e. 'Kapitza-Dirac restriction') and secondly, the incoming particles must have specified spin-polarization.

In this context, the current work theoretically investigates the high-energy Compton scattering without such restrictions, i.e. of 'unpolarized' electrons and photons with arbitrary momentum, to study whether the generation of entangled spin states of electron-photon pairs is possible under certain conditions. In this work, the term 'high-energy' refers to the order of electron energy where the rest mass energy of electron (~0.5MeV) can be neglected. For instance, the significance of rest mass energy of a 5MeV electron is ~0.5%. It is worthy to mention that the circular polarization states of a γ-photon with such order of energy can be measured experimentally [20]. Also the γ-photon absorption at such energy range belongs to the Compton scattering dominated region where neither photoelectric effect nor the electron-positron pair production is significant [21].

The two lowest possible order Feynman diagrams for such electron-photon scattering through 's'-channel and 'u'-channel are depicted in Fig.1 and the corresponding invariant amplitudes for scattering are represented by $\Pi_1$ and $\Pi_2$. Conventionally the high-energy scattering probabilities for unpolarized particles are estimated by summing over all spin states of generated particles, expecting each of them to be in superposed states since the spin states are not measured [22]. In such case, the net scattering probability is obtained to be $\langle |\Pi|^2 \rangle = \langle \Pi_1^2 + \Pi_2^2 \rangle$ since the cross-channel probability ('s'↔'u') averaged over all spin states is $\langle \Pi_1 * \Pi_2 \rangle = 0$. Thus, the generated electrons and photons are conventionally assumed to be in product state of their individually superposed states. It is apparent that, if measured, such individually superposed electron and photon out-scattered states will collapse to the corresponding specific polarization states, however, that does not necessarily imply that, when not measured, all of the outgoing electrons and photons will be in such individually superposed states and mutually in product state in general. In fact, the possibility of generation of electrons and photons of specific spin polarization on scattering of such unpolarized ones must not depend on whether such spin states are measured or not.

In this regard, the current work reveals the probability of generation of such specific spin polarized states, i.e. up/down-spin ($|\uparrow\rangle, |\downarrow\rangle$) of electron and left/right circular polarization ($|L\rangle, |R\rangle$) of photon, at the cross-channel ('s'↔'u') of high-energy Compton scattering of unpolarized electron-photon. It is worthy to mention that such pair-production at cross-channel implies a transition from one channel to the other, i.e. 's'→'u' or 'u'→'s', during scattering process. In Compton scattering, physically, the 's'-channel represents simultaneous annihilation of electron and photon whereas 'u'-channel refers to their consecutive annihilation. The propagator 4-momentum is thus different in the two channels in general (Fig.1) and equalizes to each other only under Kapitza-Dirac restriction. Therefore, without such restriction the conservation principle in general requires that any one of the two possible channel transitions

(i.e. 's'→'u' or 'u'→'s') must be accompanied together with the other one as per the limit of energy-time and position-momentum uncertainty. Such channel exchange is thus associated with a momentum transfer which in turn can induce specific polarization information to the created photon due to gauge symmetry; which in correspondence to the spin conservation induces specific spin information to the created electron. In such way it is possible to get spin/polarization entangled electron-photon pairs on Compton scattering of unpolarized electrons and photons due to channel exchange at high-energy.

Thus in general, the out-scattered state of electron-photon pair can be expressed as

$$|\psi\rangle = c_{11}|\uparrow;R\rangle + c_{12}|\uparrow;L\rangle + c_{21}|\downarrow;R\rangle + c_{22}|\downarrow;L\rangle \tag{1}$$

To quantify the entanglement between electron and photon in such case, the 'concurrence' is a good measure [23-25], given by,

$$C = 2|c_{11}c_{22} - c_{12}c_{21}| \tag{2}$$

where the maximum value of $C=1$ indicates maximally entangled pair production whereas a zero value corresponds to the product state. At this point it is worthy to mention that the cross-channel scattering probability ($\Pi_1 {}^*\Pi_2 + \Pi_2 {}^*\Pi_1$) of high energy electron/photon does not provide the phases of coefficients ($c_{ij}$) and the concurrence value directly; however, it intrinsically includes the phase factors that leads to obtain the concurrence, maximized/minimized over all possible values of such phase factors, as $C_{max/min} = 2\big||c_{11}||c_{22}| \pm |c_{12}||c_{21}|\big|$, where $(C_{max})_{min}$ or $(C_{min})_{max}$ indicates the conditions for entanglement. However, in practical experiment, for a given values of energy and angle of the in-scattering electron/photon, the out-scattered particles exhibit an angular distribution. Therefore, the aim of the current work is not only to find out the conditions for entanglement, but also to predict how much is the probability to obtain such entangled electron/photon pairs at such particular scattering angles compared to other angles, which necessitates the study of individual possibilities of getting different spin/polarization combinations.

The cross-channel ('s'↔'u') terms in such high energy Compton scattering probability are given by [22],

$$\Pi_2 * \Pi_1 = \frac{e^4}{P^2 \overline{P}^2} \left[\varepsilon_{\overline{\mu}}^k * \varepsilon_{\mu}^k \overline{\varepsilon}_{\nu}^{\overline{k}} * \overline{\varepsilon}_{\overline{\nu}}^{\overline{k}}\right] Tr\left[\left(u_{\overline{p}}^r \widetilde{u}_{\overline{p}}^r\right) \gamma^{\nu} \left[\gamma^{\alpha} P_{\alpha}\right] \gamma^{\mu} \left(u_p^s \widetilde{u}_p^s\right) \gamma^{\overline{\nu}} \left[\gamma^{\beta} \overline{P}_{\beta}\right] \gamma^{\overline{\mu}}\right] \quad (3.a)$$

$$\Pi_1 * \Pi_2 = \frac{e^4}{P^2 \overline{P}^2} \left[\varepsilon_{\overline{\mu}}^k * \varepsilon_{\mu}^k \overline{\varepsilon}_{\nu}^{\overline{k}} * \overline{\varepsilon}_{\overline{\nu}}^{\overline{k}}\right] Tr\left[\left(u_{\overline{p}}^r \widetilde{u}_{\overline{p}}^r\right) \gamma^{\mu} \left[\gamma^{\beta} \overline{P}_{\beta}\right] \gamma^{\nu} \left(u_p^s \widetilde{u}_p^s\right) \gamma^{\overline{\mu}} \left[\gamma^{\alpha} P_{\alpha}\right] \gamma^{\overline{\nu}}\right] \quad (3.b)$$

where, $(p, \overline{p}, k, \overline{k})$ represent the momentum 4-vectors of incoming and outgoing electron and photon, respectively, with $P = (p+k) = (\overline{p} + \overline{k})$ and $\overline{P} = (p - \overline{k}) = (\overline{p} - k)$; $\left(u_p^s, \varepsilon_{\mu}^k\right)$ represents Dirac spinor of spin-state '$s$' for electron and photon of polarization '$\mu$'; $\gamma$'s are Dirac gamma matrices and $e$ is the electronic charge. Since the incoming electron and photon are assumed to be unpolarized, the relations $\sum_s \left(u_p^s \widetilde{u}_p^s\right) = \gamma^{\rho} p_{\rho}$ and $\sum_{R,L} \varepsilon_{\overline{\mu}}^k * \varepsilon_{\mu}^k = -g_{\mu\overline{\mu}}$ give rise to the net 'channel-exchange' probability to be,

$$\Pi_1 * \Pi_2 + \Pi_2 * \Pi_1 = -2 \frac{e^4}{P^2 \overline{P}^2} \left[\overline{\varepsilon}_{\nu}^{\overline{k}} * \overline{\varepsilon}_{\overline{\nu}}^{\overline{k}}\right] Tr\left[\left(u_{\overline{p}}^r \widetilde{u}_{\overline{p}}^r\right) \gamma^{\rho}\right] \gamma^{\nu} \gamma^{\alpha} \gamma^{\beta} \gamma^{\overline{\nu}}\right] P_{\alpha} \overline{P}_{\beta} p_{\rho} \quad (4)$$

It is interesting to note that the generated states of electron-photon pair are non-separable due to the terms $\left(u_{\overline{p}}^r \widetilde{u}_{\overline{p}}^r\right)$ and $\left(\overline{\varepsilon}_{\nu}^{\overline{k}} * \overline{\varepsilon}_{\overline{\nu}}^{\overline{k}}\right)$ appearing together in the expression of 'channel-exchange' probability. It is also worthy to mention that prior to the calculation of such cross-channel terms to examine the possible entanglement, the net 's' and 'u'-channel scattering probability ($\langle |\Pi|^2 \rangle = \langle \Pi_1^2 + \Pi_2^2 \rangle$) is calculated and calibrated by comparing with relevant experimental results **[26]** [see **Supplementary Information**].

As shown in the schematic of the electron-photon scattering in Fig.2, without loss of generality, the momentum of outgoing electron is assumed to be along Z-axis (i.e. electron-detector position), that of the incoming electron in ZX-plane at an angle $\theta_e$ with Z-axis, and the photon to come in at the polar/azimuthal angles $(\theta_{ph}, \phi_{ph})$. Thus, in Dirac-Pauli representation, $u_{\overline{p}}^r \widetilde{u}_{\overline{p}}^r = \frac{1}{2}(I \pm \gamma^5) \gamma^{\sigma} \overline{p}_{\sigma}$, where '+' is for up-spin ($r = 1$) and '-' stands for down-spin ($r = 2$). Here, the polarization states of real photon propagating along Z-axis is given by $\overline{\varepsilon}_{R/L} \equiv 1/\sqrt{2}\,(0, \mp 1, -i, 0)$ as per right/left circular polarization (RCP/LCP), respectively, which can be transformed by using the relevant rotation matrix for any other direction of propagation. Thus, it finally results to [see **Supplementary Information**],

$$\Pi_1 * \Pi_2 + \Pi_2 * \Pi_1 = -\frac{e^4}{P^2 \bar{P}^2}\left[\bar{\varepsilon}_\nu^{\bar{k}} * \bar{\varepsilon}_{\bar{\nu}}^{\bar{k}}\right](T_1 + T_2) \qquad (5)$$

where,

$$T_1 = 8\left[(p \cdot k)\left((p-\bar{p})^\nu \bar{P}^{\bar{\nu}} + \bar{P}^\nu (p-\bar{p})^{\bar{\nu}}\right) + (p \cdot \bar{k})\left((p-\bar{p})^\nu P^{\bar{\nu}} + P^\nu (p-\bar{p})^{\bar{\nu}}\right)\right] \qquad (6.a)$$

$$T_2 = \mp 8i\varepsilon^{\nu\alpha\beta\bar{\nu}} P_\alpha \bar{P}_\beta (p \cdot \bar{p}) \qquad (6.b)$$

The '$\pm$'-sign appears in Eq.5 twice, once through $\bar{\varepsilon}_{R/L}$ for polarization of real photon and then through $T_2$ due to spin of electron, which together, in turn, results to the possible 'electron spin'-'photon polarization' entanglement.

It is observed by varying all the parameters that maximal entanglement is obtained at some specific relative directions of the inscattering electron and photon, and only if their energy values are nearly identical. In this context, at a perpendicular electron-detector position (i.e. $\theta_e = \pi/2$), the probability of producing the four possible states for different polar and azimuthal angles of the incoming photon is plotted in Fig.3. It is apparent from the figure that the probability of producing $|\uparrow;R\rangle$ is identical to that of $|\downarrow;L\rangle$, whereas the probability values for the remaining two, $|\uparrow;L\rangle$ and $|\downarrow;R\rangle$, are identical, which is obvious since the choice of 'up' or 'down' spin of electron and 'right' or 'left' circular polarization of photon is completely arbitrary. We call the former two (i.e. $|\uparrow;R\rangle$ and $|\downarrow;L\rangle$) to be 'set-1' and the latter two (i.e. $|\uparrow;L\rangle$ and $|\downarrow;R\rangle$) as 'set-2'. The plots of Fig.3 shows that, at certain angles of incoming photon direction, the nature of variation of 'set-1' and 'set-2' are opposite indicating the 'concurrence' to be high. And at a particular direction ($\theta_{ph} = \pi/2; \phi_{ph} = \pi$), 'set-1' shows a jump hike to maximum whereas 'set-2' jumps down to zero, thereby obtaining the maximum 'concurrence'. Under such specific condition, that, the incoming unpolarized electron and photon come from opposite directions with identical energy, the channel-exchange probability is obtained to be [see **Supplementary Information**],

$$\Pi_1 * \Pi_2 + \Pi_2 * \Pi_1 = 2e^4 \left[ 1 - (\pm)_{el} (\mp)_{ph} 1 \right] \tag{7}$$

where, $(\pm)_{el}$ represent up/down spin of electron and $(\mp)_{ph}$ denote RCP/LCP states of the out-scattered photon. Thus the probability of creating up-spin electron and RCP photon or down-spin electron and LCP photon is maximum and that of creating any of the other two combinations is zero indicating a maximally entangled (i.e. $C=1$), i.e. a pure Bell-state of electron-photon pair production on high energy Compton scattering due to 'channel-exchange'.

By varying the directions of incoming electron and photon, it is observed that such 'concurrence' of jumps occur at some particular values of $\theta_e$ and $\theta_{ph}$ but only at $\phi_{ph} = \pi$, i.e. when the electron-detector (i.e. the created electron) is in the plane of inscattering electron and photon. Therefore, the normalized scattering probability with all the out-scattered states $|\uparrow;R\rangle$, $|\downarrow;L\rangle$, $|\uparrow;L\rangle$ and $|\downarrow;R\rangle$ at $\phi_{ph} = \pi$ are plotted with incoming electron and photon directions $(\theta_e;\theta_{ph})$ in Fig.4 which reveals such jump-concurrence to occur only at five coordinates: $(\theta_e = \pi/4; \theta_{ph} = \pi/4)$, $(\theta_e = \pi/4; \theta_{ph} = 3\pi/4)$, $(\theta_e = 3\pi/4; \theta_{ph} = \pi/4)$, $(\theta_e = 3\pi/4; \theta_{ph} = 3\pi/4)$ and $(\theta_e = \pi/2; \theta_{ph} == \pi/2)$. However, the plots of Fig.4 show that there are many other possible directions for which 'set-1' and 'set-2' varies oppositely in nature but not as jump thereby indicating some degree of entanglement but not strong.

In order to investigate the jump-concurrence at such particular directions, the angular distribution of creation probability of 'set-1' and 'set-2' in terms of incoming photon directions for the relevant electron angles $(\theta_e = \pi/4; \pi/2; 3\pi/4)$ are plotted in Fig.5. It is to be noted that the electron-detector is at the position $\theta_{ph} = 0$ of Fig.5 and the distance of the curves from origin of the circles represent the corresponding creation probability, normalized for the respective electron angles. It is observed from the figure that at the particular angles $(\theta_e = \pi/4; \theta_{ph} = \pi/4)$, $(\theta_e = \pi/4; \theta_{ph} = 3\pi/4)$, $(\theta_e = 3\pi/4; \theta_{ph} = \pi/4)$, $(\theta_e = 3\pi/4; \theta_{ph} = 3\pi/4)$ and $(\theta_e = \pi/2; \theta_{ph} == \pi/2)$, 'set-1' and 'set-2' curves show peaks and dips, respectively. However only at $(\theta_e = \pi/2; \theta_{ph} = \pi/2)$, the variation shows a dip of 'set-2' to zero value thereby indicating the generation of a pure Bell-state. The other jump-concurrences indicate strong entanglement but not maximal. It is worthy to mention that the former four angles represent two opposite cones around the electron-detector axis and the last angle shows a plane perpendicular to it resulting maximal entanglement. It is also interesting to note that all such jump-concurrences, as can be seen in Fig.4 and Fig.5, ensure a contour line in $(\theta_e, \theta_{ph})$-space around the concurrence point on which concurrence is zero where 'set-1' and 'set-2' intersects. Thus all such strong entanglement conditions defined by jump-concurrences are bounded by the product-state conditions indicating such strong entanglement conditions to be highly selective over the scattering electron-photon directions.

In conclusion, the requirement of a quantum communication network to achieve a combination of matter qubits and propagating qubits for quantum teleportation and quantum repeater to process quantum information between remote qubits necessitates the creation of quantum entanglement between electron and photon over long distances, which is the key challenge of the state-of-the-art quantum technology [14-16]. In this context, the current work explores the possibility of entangled electron-photon pair production due to channel-exchange in high-energy Compton scattering of unpolarized electron-photon without the Kapitza-Dirac restriction. It has been shown that if the electrons and photons, with nearly equal energy of the order of MeV, scatter, an entangled pair of up-spin electron and RCP photon or down-spin electron and LCP photon can be created. Such entanglement is quite strong for particular directions of scattering particles along two opposite cones at $45^0$ around the electron-detector axis, and along the plane perpendicular to it is maximal; and thus in such particular directions the spin-polarization entangled electron-photon pairs are highly possible to be found in practical experiments. Therefore, the present work provides a protocol for creating entangled electron-photon pairs for large distance teleportation associated with future quantum communication network.

**Acknowledgement**

The authors would like to acknowledge the Centre of Excellence (COE), University of Calcutta, funded by the World Bank through TEQIP Phase III for funding the fellowship of Dr. Basudev Nag Chowdhury through the University of Calcutta.

**FIGURES**

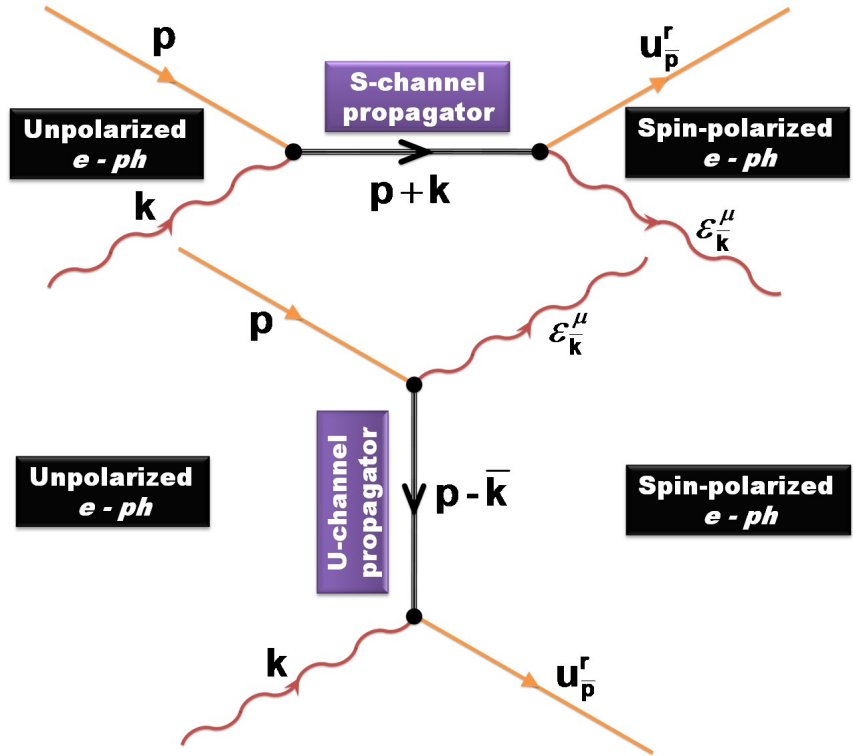

Fig.1. The two lowest possible order Feynman diagrams for Compton scattering of unpolarized electron and photon through 's'-channel and 'u'-channel resulting in the generation of spin-polarized electron-photon pair.

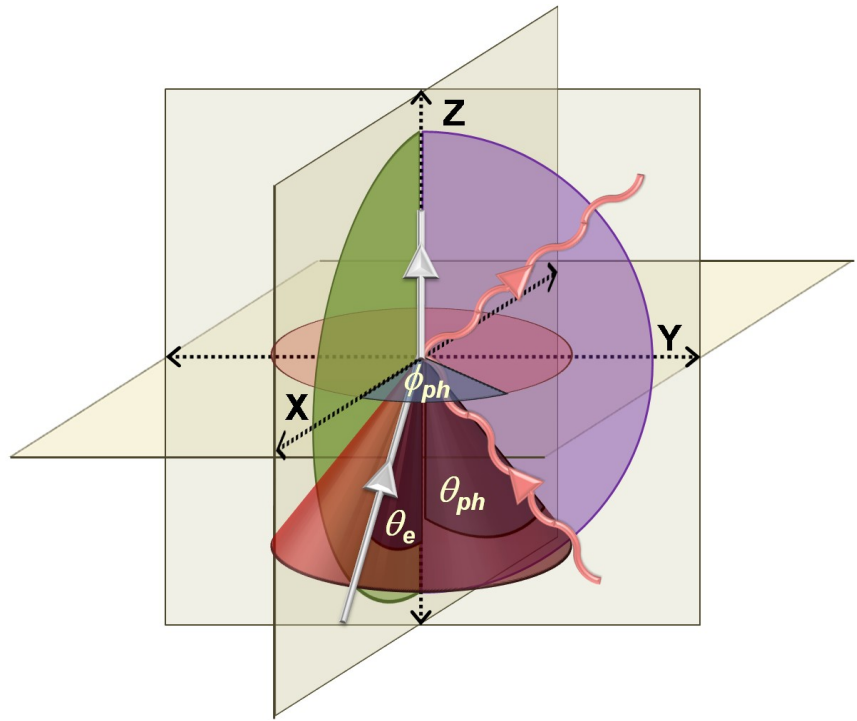

Fig.2. The schematic of scattering angles for Compton scattering of electron and photon. The outgoing electron is assumed to traverse along +ve Z-axis and the incoming electron to move on ZX-plane without loss of generality. The angle of incidence of the incoming photon is thus varied in the entire range.

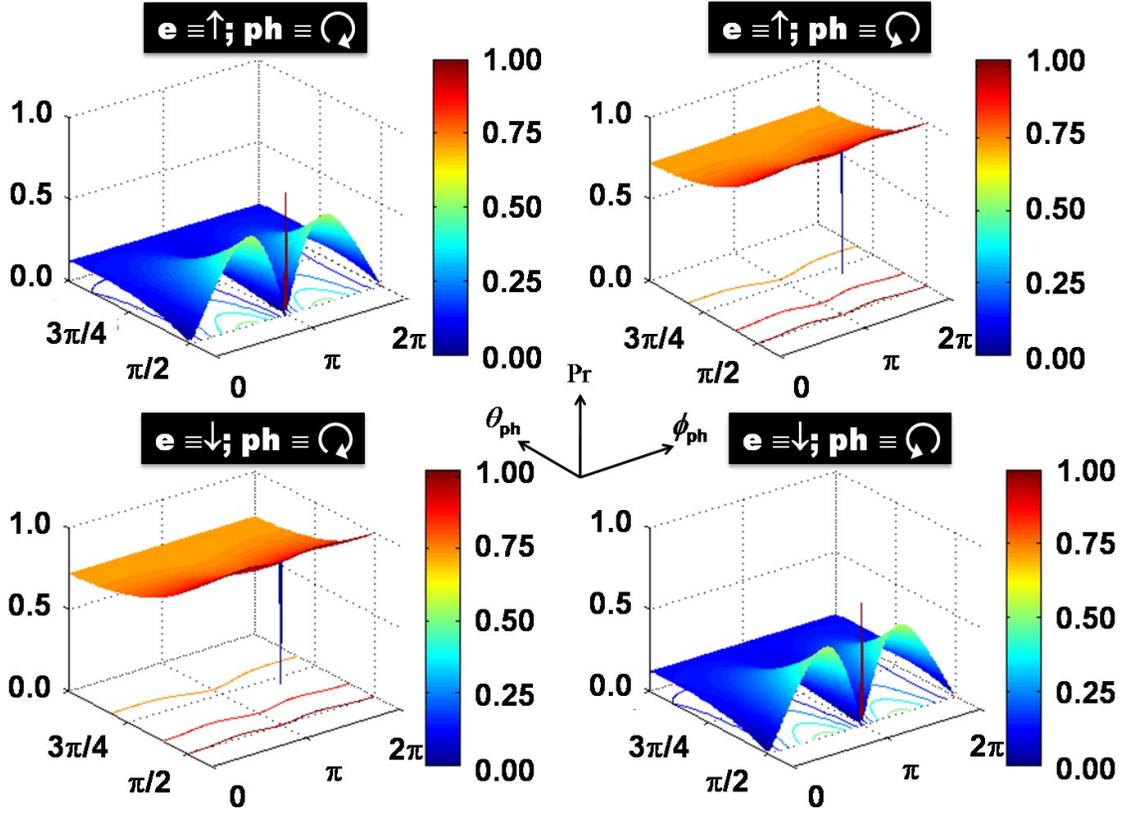

Fig.3. Normalized probability (Pr) of producing the four possible states for different polar and azimuthal angles $(\theta_{ph}, \phi_{ph})$ of the incoming photon at a perpendicular electron detector position (i.e. $\theta_e = \pi/2$).

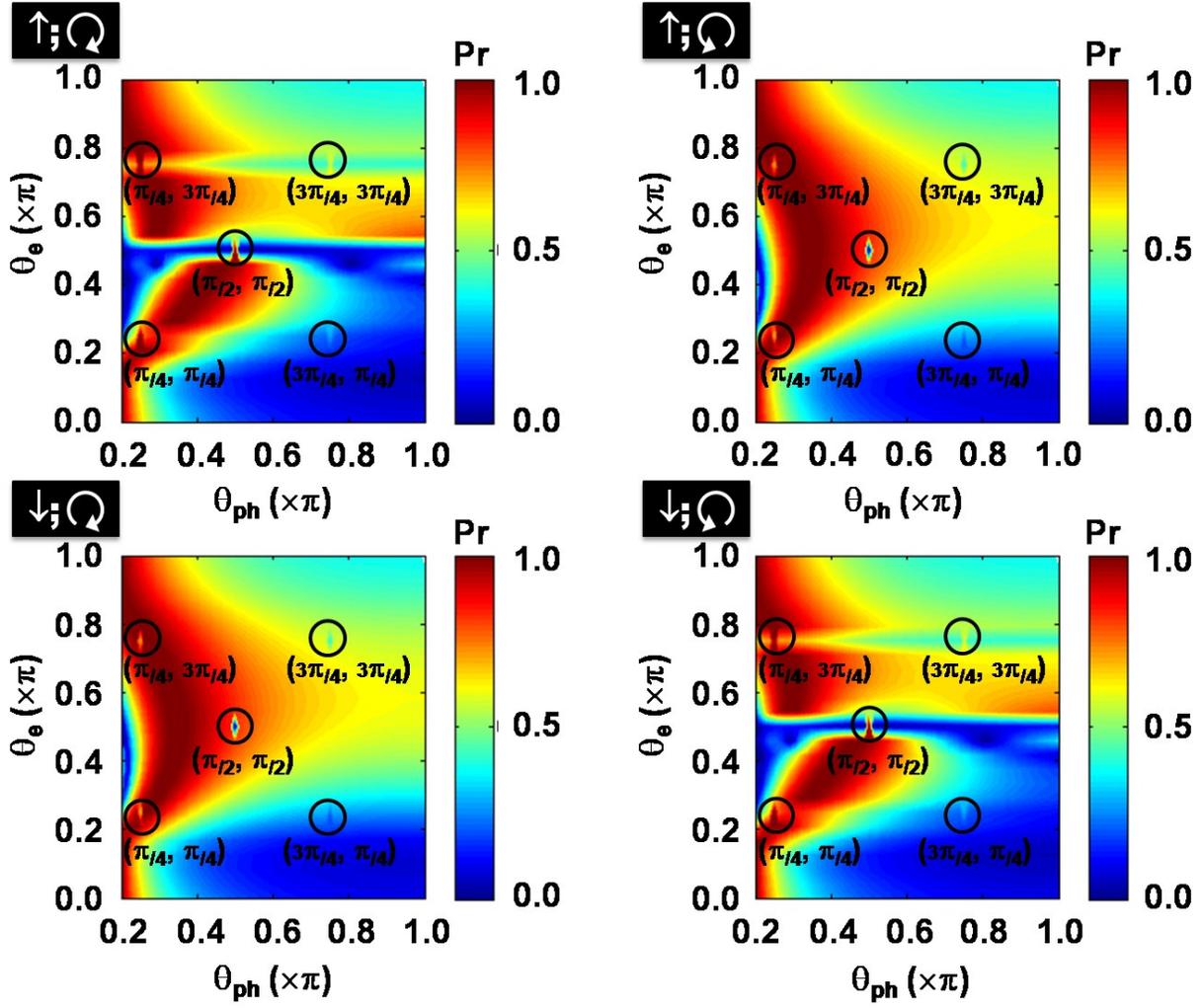

Fig.4. Normalized probability (Pr) of producing the four possible states for different polar angles $(\theta_{ph}, \phi_e)$ of the incoming coplanar electron and photon.

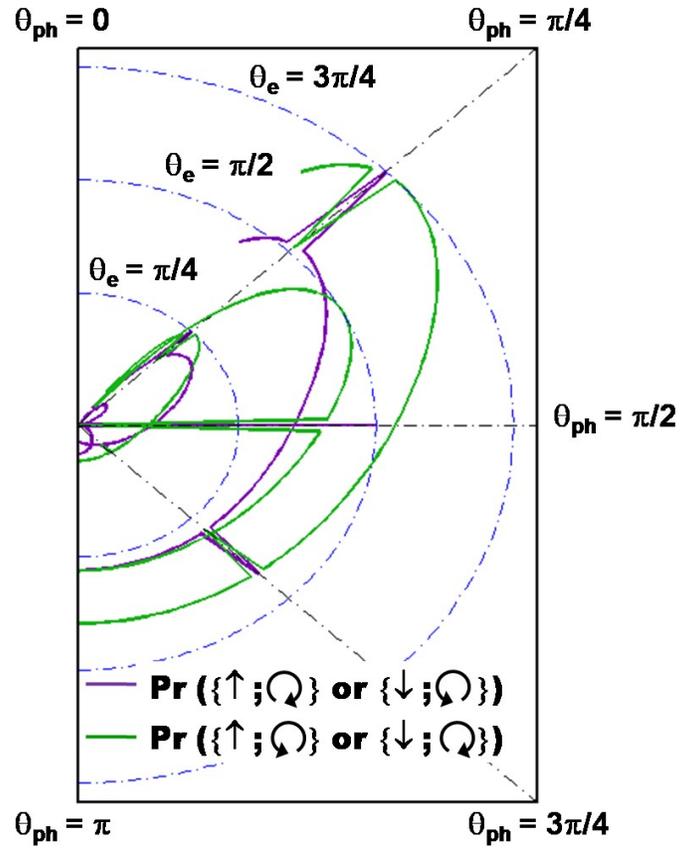

Fig.5. Angular distribution of probability (Pr) of producing the possible states over photon incidence angle ($\theta_{ph}$) for different polar angles ($\theta_e$) of the incoming electron.